\documentclass[11pt]{article}
\usepackage{epsfig}

\def\laq{~\raise 0.4ex\hbox{$<$}\kern -0.8em\lower 0.62
ex\hbox{$\sim$}~}
\def\gaq{~\raise 0.4ex\hbox{$>$}\kern -0.7em\lower 0.62
ex\hbox{$\sim$}~}

\def\beq{\begin{equation}}
\def\eeq{\end{equation}}
\def\bea{\begin{eqnarray}}
\def\eea{\end{eqnarray}}

\def \pa {\partial}

\def \ra {\rightarrow}

\def \b {\beta}
\def \a {\alpha}
\def \ap {\alpha^{\prime}}

\def \ga {\gamma}

\def \ep {\epsilon}
\def \r {\rho}
\def \om {\omega}
\def \Om {\Omega}
\def \noi {\noindent} 

\def \Ms {M_{\rm s}}
\def \Mp {M_{\rm P}}

\begin{document}
\begin{titlepage}

\begin{flushright}
BA-TH/05-521\\
CERN-PH-TH/2005-216\\
hep-th/0511039
\end{flushright}

\vspace{1.5 cm}

\begin{center}

\huge{Observable (?) cosmological signatures \\of superstrings 
in pre-big bang models of inflation}

\vspace{1cm}

\large{M. Gasperini$^{1,2,3}$ and S. Nicotri$^{2}$}

\bigskip
\normalsize

\bigskip
 $^{1}${\sl CERN, Theory Unit, Physics Department, \\ CH-1211 Geneva 23, Switzerland} 

\smallskip
$^{2}${\sl Dipartimento di Fisica, Universit\`a di Bari, \\
Via G. Amendola 173, 70126 Bari, Italy}

\smallskip
$^{3}${\sl Istituto Nazionale di Fisica Nucleare, Sezione di Bari, 
Bari, Italy}

\vspace{1.5cm}

\begin{abstract}
\noi
The different couplings of the dilaton to the $U(1)$ gauge field of heterotic and Type I superstrings may leave an imprint on the relics of the very early cosmological evolution. Working in the context of the pre-big bang scenario, we discuss the possibility of discriminating between the two models through cross-correlated observations of cosmic magnetic fields and primordial gravitational-wave backgrounds.
\end{abstract}
\end{center}

\bigskip
\bigskip
\centerline{To appear in {\bf Phys. Lett. B}}

\vfill
\begin{flushleft}
CERN-PH-TH/2005-216\\
November 2005 
\end{flushleft}

\end{titlepage}

\newpage
\parskip 0.2cm

It is well known that the five superstring models differ, at the level of the lowest-order bosonic effective action, for their field content and for the way in which the various fields are coupled to the dilaton \cite{1}. Even in models with the same (boson) particle content, such as Type I and heterotic superstrings, there are differences due to the dilaton couplings, most notably for the Yang--Mills one form containing the $U(1)$ gauge field associated to standard electromagnetic (e.m.) interactions. 

The effective tree-level action for the photon--dilaton interactions, after appropriate dimensional reduction from ten to four dimensions, can be parametrized (in the string frame) as follows:
\beq
S=-{1\over 4} \int d^4 x \sqrt{-g} \, f(b_i)\,e^{-\ep \phi}\, F_{\mu\nu}F^{\mu\nu},
~~~~~~~~~~F_{\mu\nu}= \pa_\mu A_\nu-\pa_\nu A_\mu,  
\label{1}
\eeq
where $\phi$ is the ten-dimensional dilaton, (normalized in such a way that the string coupling parameter $g$ corresponds to $g^2= \exp \phi$), and where the ``form factor" $f(b_i)$ takes into account the possibe dynamical contribution of the internal moduli fields $b_i$, $i=1, \dots , 6$. 
The dilaton parameter $\ep$ has values $1, 1/2$, respectively, for heterotic and Type I superstrings, 
which are probably the best candidates for representing the low energy limit of standard model interactions. 
The value of $\ep$ could be different, however, were the e.m. field identified not as a component of the higher-dimensional gauge group, but  as a one form present in the Ramond--Ramond or  Neveu-Schwarz--Neveu-Schwarz sector of other string theory models. In that case also the coupling to the internal moduli would be different (see for instance \cite{22}). 

The discussion of this paper will be concentrated on the case in which $f(b_i)$ simply represents the volume factor of the six-dimensional internal manifold, $f(b_i)= V_6 \equiv \prod_{i=1}^6  b_i$, and different string models are characterized by different values of $\ep$. When the internal moduli are stabilized, indeed, the coupling function $f(b_i)$ becomes trivial, and the only relevant parameter is $\ep$. In any case, different values of $\ep$ amount to different rescaling of the photon--dilaton coupling, and may have significant impact on cosmological processes where large variations of the dilaton field, for long periods of time, may occur. 

The effect we shall consider here will concern the amplification of the quantum fluctuations of the e.m. field, and the possible production of ``seeds" for the magnetic fields currently observed on large scales \cite{2}. Working in the context of the so-called pre-big bang scenario of superstring cosmology \cite{2a,5} it will be shown in particular that, depending on the value of $\ep$, the region of parameter space compatible with an efficient production of seeds {\em may} or {\em may not} overlap with the different regions of parameter space compatible with an observable production of cosmic gravitons. The cross-correlated analysis of cosmic magnetic fields and of the outputs of gravitational detectors may thus provide unique information, in principle, not only on the allowed region of parameter space of the considered class of string cosmology models, but also on the most appropriate superstring model -- i.e. on the most appropriate M-theory limit -- to be chosen for a realistic description of fundamental interactions during the primordial stages of our Universe. 

We start by recalling that the e.m. fluctuations of the vacuum, according to the action (\ref{1}), are represented by the canonical variable $\psi_\mu$,
\beq
\psi_\mu =z_\ga A_\mu, ~~~~~~~~~~~~
z_\ga =\left( \prod_{i=1}^6  b_i\right)^{1/2} e^{-\ep \phi/2}, 
\label{1a}
\eeq
which diagonalizes the four-dimensional action.  In a conformally flat four-dimensional geometry, $g_{\mu\nu}=a^2 \eta_{\mu\nu}$, and in the radiation gauge $A_0=0=\pa_iA_i$, $\psi_\mu$ satisfies the canonical equation \cite{3}:
\beq
\psi_i''- \nabla^2 \psi_i - z_\ga^{-1}z_\ga''\psi_i=0,
\label{2}
\eeq
where the  primes denote differentiation with respect to the conformal time $\eta$. Notice that, because of the conformal invariance of the four-dimensional Maxwell action, the fluctuations $\psi_i$ are decoupled from the four-dimensional geometry; the canonical equation only contains a coupling to the first and second derivatives of the dilaton background and of the internal moduli. A time-dependent dilaton (with the possible contribution of a time-dependent internal geometry) may thus act as an external ``pump field", and sustain the amplification of the e.m. fluctuations during the primordial phases of accelerated (inflationary) evolution. 

For the case of the heterotic coupling $\ep=1$ the inflationary spectrum of fluctuations produced by the dilaton has already been computed in \cite{3,4}. Here we shall derive the corresponding spectrum for a generic value of $\ep$, using the same example of inflationary string-cosmology background as was adopted in \cite{3},  based on a ``minimal" pre-big bang scenario consisting of two phases.

During the first phase, ranging in conformal time from $-\infty$ to $\eta_s<0$, the Universe is assumed to evolve from the string perturbative vacuum according to the exact (Kasner-like) solutions of the lowest order gravi-dilaton effective action \cite{5}: 
\beq
a(\eta)\sim (-\eta)^{\b_0\over (1-\b_0)}, ~~~~
b_i (\eta) \sim (-\eta)^{\b_i\over(1-\b_0)}, ~~~~
\phi(\eta) \sim {\sum_i \b_i +  3 \b_0-1\over 1-\b_0} \ln (-\eta), ~~~~
\eta<\eta_s <0,
\label{3}
\eeq 
present in all superstring models. Here $a$ is the (string-frame) scale factor of the four-dimensional, isotropically expanding geometry, and the parameters $\b_0, \b_i$ are related by the Kasner condition
\beq
3 \b_0^2+ \sum_i \b_i^2=1.
\label{3a}
\eeq
The evolution of the pump field $z_\ga$, in this first phase, is controlled by $\b_0, \b_i$ as follows:
\beq
z_\ga (\eta) \sim (-\eta)^{\ep'\sqrt 3 /2}, ~~~~~~~~~~~
\ep'= {\sum_i\b_i -\ep(\sum_i\b_i+3\b_0-1)\over \sqrt 3 (1-\b_0)}
\label{3b}
\eeq
(in the special case of frozen internal dimensions, $\b_i=0$, $\b_0=-1/\sqrt{3}$ and $\ep'=\ep$). In the following discussion we will consider, for simplicity, an isotropic internal geometry, so that the parameter $\b_i$ can be eliminated everywhere in terms of $\b_0$ through the Kasner condition. We will also assume -- for a realistic inflationary scenario evolving from the higher-dimensional perturbative vacuum towards the four-dimensional strong coupling regime -- that the internal dimensions are shrinking, and that the four-dimensional coupling, controlled by $\phi_4=\phi-\ln V_6=\phi-6 \ln b_i$, is growing. The intersection with the Kasner condition then defines, for $\b_0$ and $\b_i$, the following allowed ranges of values:
\beq
-1/\sqrt3 \leq \b_0 <1/3, ~~~~~~~~~~~~~~
0< \b_i \leq 1/\sqrt 6.
\label{3c}
\eeq

During the second, high-curvature phase, ranging from $\eta_s$ to $\eta_1<0$, the (string-frame) curvature stays frozen at a constant value $H_1$ controlled by the string scale, $H_1 \sim \Ms$; the internal moduli are also frozen, while the dilaton (and the string coupling $g$) keeps growing from the initial value $g_s^2 = \exp \phi(\eta_s) \ll1$ to a final value $g_1^2 = \exp \phi(\eta_1) \sim 1$. The background evolution during this second ``stringy" phase can be represented by
\beq
a(\eta)\sim (-\eta)^{-1}, ~~~
\phi(\eta) \sim -2 \b \ln (-\eta), ~~~ z_\ga(\eta) \sim (-\eta)^{\ep \b}, 
 ~~~~\b >0, ~~~\eta_s \leq \eta \leq \eta_1 <0,
\label{4}
\eeq
where $\b$ is a phenomenological parameter that possibly takes into account the effects of the higher order $\ap$ corrections to the effective action. Eventually, for $\eta>\eta_1$, the Universe is expected to enter the radiation-dominated, frozen-dilaton (decelerated) regime $a \sim \eta$, $\phi=$ const, and to follow all subsequent steps of the standard cosmological evolution. 

For such a model of background, each polarization component of the 
canonical variable can be expanded in Fourier modes $\psi_k$ such that $\nabla^2 \psi_k=-k^2 \psi_k$, and Eq. (\ref{2}) becomes a Bessel equation whose solution can be expressed in terms of the first- and second-kind Hankel functions \cite{5}. By imposing the asymptotic normalization to a spectrum of quantum vacuum fluctuations, $\psi_k= \exp(-ik\eta)/\sqrt{2k}$ for $\eta \ra -\infty$, the solution is then fully specified by the matching conditions of $\psi$ and $\psi'$ at $\eta=\eta_s$ and $\eta=\eta_1$. Since there are two background transitions, the final spectral distribution for $\psi_k(\eta)$, $\eta>\eta_1$ will contain in general two branches: a high-frequency branch, for modes $k>k_s=|\eta_s|^{-1}$ that are only affected by the last transition at $\eta_1$, and a low-frequency branch, for modes $k<k_s$ that are affected by both transitions. For $k>k_1=|\eta_1|^{-1}$ the amplification (or particle production) is exponentially suppressed as $\exp(-k/k_1)$ \cite{6}, and can be consistently neglected within the level of approximation adopted in this paper. 

Switching to proper frequencies, $\om=k/a$, a straightforward computation of the final spectral energy density of the amplified  fluctuations, expressed in critical units, and referred to the present epoch $t_0$, leads to
\bea
\Om_\ga(\om, t_0) \equiv {d (\r_\ga/\r_c)\over d \ln \om}
&=& \Om_r(t_0)\left(H_1\over M_{\rm P}\right)^2\left(\om \over \om_1\right)^{3-|2\ep\b-1|},   
~~~~~~~~~~~~~~~~~~~~~~~~~\om_s< \om <\om_1,
\nonumber\\
&=&
 \Om_r(t_0)\left(H_1\over M_{\rm P}\right)^2\left(\om_s \over \om_1\right)^{3-|2\ep\b-1|} \left(\om\over \om_s\right)^{3-|\ep'\sqrt 3-1|},
~~~~~\om<\om_s, 
\nonumber\\ &&
\label{5}
\eea
where the parameters $\b, \b_i, \b_0$ may vary in the ranges previously defined, and where we have assumed $\ep>0$ (for $\ep=0$ there is no amplification of e.m. fluctuations during the string phase, and the above spectrum cannot be applied). 
HereΠ$\r_c=3 H_0 \Mp^2$ is the critical energy density, $\Mp= (8 \pi G)^{-1}$ is the Planck mass, $\Om_r(t_0)\simeq 4 \times 10^{-5}h_0^{-2}$ is the present fraction of critical energy density in radiation, with $h_0=H_0/(100~ {\rm km}~ {\rm sec}^{-1} {\rm Mpc}^{-1})$ the present (dimensionless) value of the Hubble parameter. Finally, $\om_1= (a|\eta_1|)^{-1} \simeq H_1(a_1/a)$ 
and $\om_s= (a|\eta_s|)^{-1} \simeq H_s(a_s/a)$, where $H_1\sim \Ms$ is the value of the curvature scale at the inflation $\ra$ radiation transition, while $\om_s$ is a free parameter of the given inflationary scenario. Previous detailed computations of the above spectrum for the pure heterotic case $\ep=1$ can be found in \cite{5,7} (see \cite{8} for a recent computation of the spectrum for a generic value of $\ep$, but for the special case of frozen internal dimensions, i.e. with $\ep'=\ep$). 

This primordial spectrum of e.m. fluctuations is distributed over a wide range of frequencies and can provide, in principle, the ``seeds" required to trigger the galactic dynamo and to produce the cosmic magnetic fields $B_G$ (of microgauss strength) currently observed on the galactic scale \cite{2,3}. The identification of the seed fields with the vacuum fluctuations, amplified by inflation, requires however that such fluctuations be coherent and large enough over a proper length scale that today roughly corresponds to the megaparsec scale, as first pointed out in \cite{9}. 

For a conservative estimate of the required field strength \cite{9} one can assume the existence of the standard galactic dynamo mechanism, operating since the epoch of structure formation, characterized by an amplification factor $\sim 10^{13}$, and can also take into account the additional amplification  ($\sim 10^4$) due to magnetic-flux conservation in the collapse of the galactic structure from the Mpc to the $10$ kpc scale. One obtains, in this way, the lower bound $B_s \gaq 10^{-23}$ gauss on the present amplitude of the magnetic seeds at the Mpc scale; the identification of the seeds with the inflationary spectrum of e.m. fluctuations leads then to the condition \cite{9}:
\beq
{B_s^2(\om_G,t_0)\over B_G^2(t_0)} \simeq
{\r_\ga (\om_G,t_0)\over \r_r (t_0)} = {\Om_\ga (\om_G,t_0)\over \Om_r (t_0)} \gaq 10^{-34}, ~~~~
\om_G = (1 ~{\rm Mpc})^{-1} \sim 10^{-14} ~{\rm Hz}.
\label{6}
\eeq
We have used the approximate equality of the energy density associated to the galactic magnetic field ($B_G \sim 10^{-6}$ gauss) and of the energy density of the CMB radiation. 

This lower bound on $\Om_\ga$ has to be complemented by a competing upper bound, since the energy density stored into the amplified fluctuations cannot be too large to avoid destroying the large-scale homogeneity of the cosmological background, and to be consistent with the linearized treatment of the fluctuations as small perturbations, with negligible backreaction. This imposes the stringent, model-independent constraint \cite{3,7}
\beq
\Om_\ga (\om,t) \leq \Om_r(t),
\label{7}
\eeq
to be satisfied at all times, for all frequency scales of the amplified spectrum. 

The intersection of the two constraints (\ref{6}) and (\ref{7}) determines, for any given value of $\ep$, an allowed region in the parameter space of our model of background. The space is spanned by the variables $\{\om_1, \om_s, \b, \b_0, H_1\}$, and the purpose of this paper is to compare such regions with analogous allowed regions associated to the amplification of tensor metric perturbations, and to the detectable production of cosmic gravitons, in the same given class of pre-big bang backgrounds. 

To this purpose we must consider the evolution equation of the canonical variable $u_{ij}$, associated to the tensor perturbations $h_{ij}$ of the three-dimensional spatial sections of the metric background:
\beq
u_{ij}= z_g\, h_{ij}, ~~~~~~~~~~~
z_g =a\left( \prod_{i=1}^6  b_i\right)^{1/2} e^{- \phi/2}.  
\label{7a}
\eeq
In the transverse, traceless gauge $\pa_ih_j^i=0=h_i^i$, and in the string frame, such a canonical equation reads \cite{10}
\beq
u_i^{\prime \prime j}- \nabla^2 u_i\,^j -z_g^{-1}  
z_g'' u_i\,^j =0,
\label{8}
\eeq
for each spin-two polarization mode $u_i\,^j$. We will use the same model of background specified by Eqs. (\ref{3}), (\ref{4}), and we can notice that the graviton ``pump field" $z_g$, at low energies, is completely insensitive to the dynamics of the internal dimensions \cite{11,12}, unlike e.m. fluctuations: according to Eq. (\ref{3}), indeed, 
 $z_g \sim (-\eta)^{1/2}$ during the initial low-curvature phase, quite independently of the values of $\b_0$ and $\b_i$ (and even in the anisotropic case). In the second, high-curvature phase, the pump field is parametrized instead by  $z_g\sim (-\eta)^{\b-1}$, with $\b>0$, according to Eq. (\ref{4}). 

We can now follow the same procedure as before, expanding the fluctuations in Fourier modes, imposing the canonical normalization at $\eta \ra -\infty$, and matching the solutions at $\eta_s$ and $\eta_1$. The computation of the spectral energy density stored in the amplified gravitational radiation leads to the final (already known \cite{12,12a,5}) result: 
\bea
\Om_g(\om, t_0) 
&=& \Om_r(t_0)\left(H_1\over M_{\rm P}\right)^2\left(\om \over \om_1\right)^{3-|2\b-3|},   
~~~~~~~~~~~~~~~~~~\om_s< \om <\om_1,
\nonumber\\
&=&
 \Om_r(t_0)\left(H_1\over M_{\rm P}\right)^2\left(\om_s \over \om_1\right)^{3-|2\b-3|} \left(\om\over \om_s\right)^{3},
~~~~~~~~\om_{\rm eq}<\om<\om_s. 
\label{9}
\eea
We have neglected small logarithmical corrections to the low-energy branch of the spectrum (which are irrelevant for the purpose of this paper), and we have omitted the ultra-low frequency branch $\om<\om_{\rm eq}$, whose amplitude is so depressed as to be safely regarded as negligible. 

It is important, for the subsequent discussion, to note that both branches of the spectrum are growing, or at least non-decreasing. The high-frequency, string branch of the spectrum could be decreasing, in principle, for $\b<0$ or $\b>3$: the first possibility is excluded, however, for the consistency of the assumed growing-dilaton scenario (see Eq. (\ref{4})), while the second possibility is excluded since it would correspond to  too fast an evolution of the graviton pump field, $z_g \sim (\eta)^\a$, with a power $\a=\b-1>2$, which is known to lead to the gravitational instability of the background \cite{13}. We shall thus restrict our discussion to the case $0\leq \b\leq 3$. We should warn the reader, however, that the considered graviton spectrum is typical of a stringy phase at constant curvature and linearly growing dilaton \cite{15a}, which takes into account higher-derivative corrections but remains at the tree-level in the string coupling: the spectrum could be significantly different if higher-loop corrections were included. 

There are two main upper bounds on the amplitude of such a growing spectrum: one is obtained from the observations of millisecond pulsars (in particular, from the absence of a detectable distortion of pulsar timing), which imposes the bound \cite{14}
\beq
h_0^2 \,\Om_g(\om_p, t_0) \laq 10^{-8}, ~~~~~~~~~~~~~~~~
\om_p \simeq 10^{-8}~ {\rm Hz};
\label{10}
\eeq
the other is obtained from the standard nucleosynthesis analysis, which imposes a bound on the total integrated energy density of the relic graviton spectrum at the nucleosynthesis epoch \cite{15}. Such a bound can also be expressed as an upper limit on the peak intensity of the spectrum \cite{16}: 
\beq
\Om_g(\om, t_0) \laq 10^{-1}\Om_r(t_0),
\label{11}
\eeq
which is valid at all frequency scales. These two upper bounds are in competition with the lower bounds to be imposed on the spectrum for its possible detection by a gravitational antenna of given sensitivity. Here we will use, as a typical reference value, the planned sensitivity of the Advanced LIGO interferometers, which are expected to detect a background of relic gravitons through cross-correlated observations around the frequency band $\om_L \sim 10^2$ Hz, provided \cite{17}:
\beq
h_0^2\, \Om_g(\om_L,t_0) \gaq 10^{-11}, ~~~~~~~~~~~~~~~~
\om_L/2\pi \simeq 10^{2}~ {\rm Hz}.
\label{12}
\eeq
See \cite{18,5} for other sensitivities associated to different types of gravitational detectors. 

The set of constraints (\ref{10})--({\ref{12}), imposed on the graviton spectrum (\ref{9}), defines allowed regions in the parameter space of our class of background, which we shall now compare with the e.m. regions specified by the constraints (\ref{6}), (\ref{7}). For a  clearer  discussion we will first reduce the dimensions of the parameter space by exploiting the direct relationship existing between $H_1$ and $\om_1$, determined by the details of the post-inflationary evolution: considering the standard radiation-dominated $\ra$ matter-dominated evolution we find \cite{5}, in particular, $\om_1 \simeq H_1 a_1/a \simeq (H_1/\Mp)^{1/2} 10^{11}$ Hz. Next, we will fix the transition scale $H_1$ precisely at the present value of the string scale, $H_1=\Ms$, assuming for $\Ms$ the typical value $\Ms=0.1 \Mp$, consistently with string models of unified gravitational and gauge interactions \cite{19}. This completely specifies the position and the ``end point" of the graviton and photon spectrum in terms of $\om_1$ and $\Om(\om_1)$, and leaves us with four parameters, $\om_s, \b, \b_0$ and $\ep$. The graviton spectrum, however, only depends on $\om_s$ and $\b$: we will thus plot the allowed regions in the two-dimensional space spanned by the variables $\{\om_s, \b\}$, at different, constant values of the parameters $\ep$ and $\b_0$, chosen appropriately for the e.m. spectrum. 

For the physical interpretation of the final result, however, it will be convenient to replace $\om_s$ and $\b$ with an equivalent set of parameters, more directly related to the inflationary kinematics. This new set is obtained by noting that, according to Eq. (\ref{4}), $\om_s$ is related to the  extension in time of the string phase through the e-folding factor $z_s=a_1/a_s=\eta_s/\eta_1= k_1/k_s= \om_1/\om_s$; the parameter $\b$ is related instead to the growth of the string coupling during the string phase, since $g_s/g_1= (\eta_1/\eta_s)^\b= z_s^{-\b}$. On the other hand, once we have imposed $H_1=\Ms= 0.1 \Mp$, not only $\om_1$ but also the final string coupling $g_1=\exp(\phi_1/2)$ is fixed \cite{16}, at a value quite close to the currently expected value $g_1\simeq \Ms/\Mp =0.1$ (since the dilaton is assumed to be frozen during the subsequent standard evolution). The two variables $\{z_s, g_s\}$ thus represent a complete and independent set of variables, equivalent to the set $\{\om_s, \b\}$, for a convenient parametrization of the chosen class of string cosmology backgrounds \cite{3}. For practical purposes we shall work, in particular, with a decimal logarithmic scale, and we shall plot the figures in the plane parametrized by the coordinates
\beq
x=\log z_s= \log(\om_1/\om_s)>0, ~~~~~~~~~~~~
y=\log (g_s/g_1)=-\b \log z_s =- \b x<0.
\label{13}
\eeq

We are now in the position of discussing the possible overlap of the allowed regions for the photon and graviton spectra. We rewrite these  in terms of the new variables $\{x, y\}$ and we note, first of all, that the condition (\ref{11}) is automatically satisfied because of the choice $\Ms/\Mp=0.1$ and of the monotonic growth of the graviton spectrum, which imposes $0 \leq \b\leq 3$, and which translates into the stringent constraint
\beq
-3x <y<0
\label{14}
\eeq
on the allowed  gravitational region. 

A similar argument can be applied to the e.m. spectrum (\ref{5}) by noting that its dilaton (low-frequency) branch, controlled by the parameter $\ep'$, is always growing for all values of $\b_0,\b_i$ included in the allowed range (\ref{3c}), and for $\ep$ varying from $0$ to $1$. The 
high-frequency branch of the spectrum might be decreasing, in principle, if $y<-(2x/\ep)$: a decreasing spectrum, however, has a peak at $\om=\om_s$, and is only marginally compatible with the homogeneity condition (\ref{7}), which imposes, when applied to the peak value of the spectrum, $y>-(2x/\ep)-1/\ep$. In addition, a decreasing spectrum is ruled out if one would take into account the slightly more stringent (but model-dependent) bound $\Om_\ga <0.1\Om_r$, which follows from the presence of strong magnetic fields at the nucleosynthesis epoch \cite{20}, which would impose $y>-(2x/\ep)$. As in the graviton case, we will thus restrict our discussion to a growing spectrum of e.m. fluctuations, which automatically satisfies Eq. (\ref{7}) and the nucleosynthesis bound, and which imposes the constraint
\beq
-2x/\ep <y<0
\label{15}
\eeq
on the allowed e.m. region. 

The allowed regions of the plane $\{x, y\}$ determined by the set of conditions (\ref{6}) (efficient seed production), (\ref{10}) (pulsar bound), (\ref{12}) (graviton detectability), (\ref{14}) and (\ref{15}) (growing spectra) are illustrated in Fig. \ref{f1}, which presents in a compact form the main results of this paper. The regions marked by the thin border refer to the graviton spectrum (\ref{9}), those marked by the bold border refer to the e.m. spectrum (\ref{5}). We have plotted, for the e.m. spectrum, the two cases $\ep=1$ and $\ep=1/2$, and we have explicitly indicated the application of the constraints to the high-frequency (string) or low-frequency (dilaton) branches of the spectrum, both for tensor and e.m. fluctuations. 

It should be stressed that the existence of two graviton regions is due to the absolute value appearing in the power of the graviton spectrum, and to the different kinematic behaviour of tensor perturbations outside the horizon \cite{12,7}, depending on the sign of $2\b-3$: the upper part of the allowed region corresponds to the class of backgrounds with pump field characterized by the power $\b-1<1/2$, in which the amplitude of tensor perturbations stays constant outside the horizon; the lower part corresponds to backgrounds with  $\b-1>1/2$, in which the amplitude grows outside the horizon, and the inflationary amplification is even more effective \cite{5}. No such distinction exists,  instead, for the allowed regions of the photon spectrum, once the condition (\ref{15}) has been imposed. 

Finally, the dashed lines present in the dilaton sector of the photon  regions illustrate the possible effects of a non-trivial (primordial) dynamics of the internal dimensions, possibly producing different values of the effective parameter $\b_0$ according to Eqs. (\ref{3}), (\ref{3a}). The bold line represents the seed condition imposed on the photon spectrum computed in the case of frozen internal dimensions, and corresponding to $\b_0=-1/\sqrt3$, $\b_i=0$; the three dashed lines are computed instead for a non-trivial internal dynamics, corresponding to three different pairs of values of the parameters $(\b_0, \b_i$). We have chosen, respectively,  $(-1/3, 1/3)$, $(0, 1/\sqrt 6)$, $(1/3,1/3)$ for the case $\ep=1$, and  $(-0.568, 0.072)$, $(-1/3, 1/3)$, $(0.063,0.406)$ for the case $\ep=1/2$ (the numerical values have been selected so as to include the maximal and minimal extension of the allowed region compatible with the range of variations of the parameters $\b_0$ and $\b_i$, given in Eq. (\ref{3c})). Note that we have also included  the possibility of a ``bouncing" scenario with positive $\b_0$, in which the initial growth of the dilaton, at low energy, is associated with the  shrinking of the four-dimensional geometry. The effects of the internal dynamics may enhance or reduce the photon allowed regions but, as clearly illustrated in the picture, it cannot lead to the overlap of the photon and graviton regions in the case $\ep=1$. 

\begin{figure}[t]
\centerline{\epsfig{file=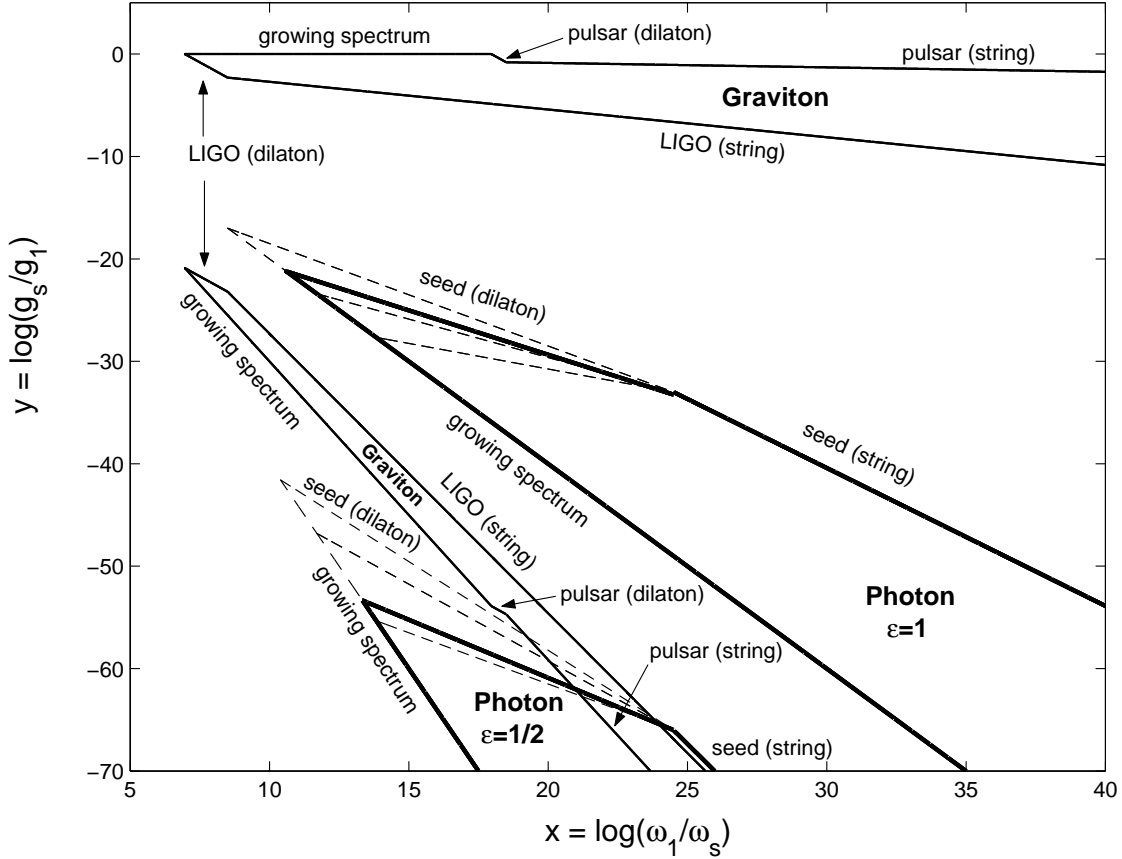, width=150mm}}
\caption{\sl Allowed regions determined by the conditions (\ref{6}), 
(\ref{10}), (\ref{12}), (\ref{14}), (\ref{15}), imposed on the spectra (\ref{5}) and (\ref{9}) with $(H_1/\Mp)=0.1$. The photon regions are plotted for the two cases $\ep=1$ and $\ep=1/2$, and for appropriate values of $\b_0$ and $\b_i$ illustrating the maximal and minimal extension of the allowed region in the case of a non-trivial evolution of the moduli fields (dashed lines).} 
\label{f1}
\end{figure}

The main message of Fig. \ref{f1} is that an efficient production of seeds for the cosmic magnetic fields  (amplified by the dilaton in a string cosmology context) is in principle compatible with the associated production of a cosmic graviton background detectable  by Advanced LIGO if $\ep=1/2$ (as in the case of Type I superstrings), and incompatible if $\ep=1$ (as in the case of heterotic superstrings). This may give us direct experimental information on the possible primordial strength of the photon--dilaton coupling, and on the choice of the superstring model most appropriate to the early cosmological evolution. 

Let us suppose, indeed, that further studies of cosmic magnetic fields will provide direct independent confirmation of the expected seed production, exactly as predicted in the class of pre-big bang models considered here: for instance, by measuring the primordial spectrum through its effects on the polarization and the anisotropy of the CMB radiation \cite{2} (primordial seeds can be produced, more or less naturally, with various mechanisms \cite{8}, but a monotonically growing spectrum seems to be peculiar only of the amplification of the vacuum fluctuations in pre-big bang cosmology). Then,  a future detection of cosmic gravitons by the next generation of gravitational antennas will provide support in favour of Type I models, while the absence of detection (at the same sensitivity level) should  be interpreted more in favour of the heterotic  model of coupling (see \cite{21} for a very recent, direct experimental  bound on the energy density of a cosmic graviton background, and \cite{21a} for  a discussion  of the near-future sensitivities of LIGO in various regions  of the parameter space of pre-big bang models). 
Such an argument cannot be directly applied  to cosmological models based on Type II superstrings, which are expected to describe the strong coupling limit of standard interactions. 

We may notice, as a conclusive remark, that the extension of the low-energy (dilaton) part of the allowed regions illustrated in Fig. \ref{f1} is rather strongly dependent on the values of $\b_0$ and $\b_i$. A precise experimental determination of the spectrum of e.m. fluctuations may thus, in principle, open a direct window on the primordial dynamics of the extra dimensions. Work along these lines is currently in progress (see also \cite{22,21b} for previous studies on higher-dimensional modifications of the photon spectrum).

\section*{Acknowledgements}
It is a pleasure to thank Carlo Angelantonj,  Alessandra Buonanno, Massimo Giovannini, Augusto Sagnotti e Gabriele Veneziano for helfpful discussions and comments.

\end{document}